\documentclass[a4paper,12pt]{article}
\usepackage{jheppub}
\usepackage{amssymb}
\usepackage{graphicx}
\usepackage{amsmath}
\usepackage{hyperref}
\usepackage{tikz}
\usepackage{bm}
\usepackage{exscale}
\usepackage{relsize}
\usepackage{placeins}
\allowdisplaybreaks[4]

\usetikzlibrary{arrows.meta,decorations.pathmorphing,backgrounds,positioning,fit,petri}
\tikzset{zigzag/.style={decorate, decoration=zigzag}}

\definecolor{darkgreen}{rgb}{0.,0.6,0.}

\newcommand{\md}{\mathrm{d}}
\newcommand{\me}{\mathrm{e}}

\title{\boldmath Vacuum bubbles from cosmic ripples}

\author[1]{Zi-Yan Yuwen,}
\emailAdd{ziyan.yuwen@apctp.org}
\author[3,4]{Rong-Gen Cai,}
\emailAdd{caironggen@nbu.edu.cn}
\author[*,2,1]{Shao-Jiang Wang}
\emailAdd{schwang@itp.ac.cn (corresponding author)}

\affiliation[1]{Asia Pacific Center for Theoretical Physics (APCTP), Pohang 37673, Korea}
\affiliation[2]{Institute of Theoretical Physics, Chinese Academy of Sciences, Beijing 100190, China}
\affiliation[3]{Institute of Fundamental Physics and Quantum Technology \& School of Physical Science and Technology, Ningbo University, Ningbo 315211, China}
\affiliation[4]{School of Fundamental Physics and Mathematical Sciences, Hangzhou Institute for Advanced Study (HIAS), University of Chinese Academy of Sciences (UCAS), Hangzhou 310024, China}

\abstract{We investigate vacuum decays in the early Universe in the presence of curvature perturbations. For sufficiently large perturbations associated with over-densities, we find that the bounce solution develops an oscillating middle stage near the bubble wall. For small perturbations, we analytically show, within the thin-wall approximation, that an over- (under-) density would enhance (suppress) the vacuum decay rate for a smaller (larger) initial bubble radius. By numerically solving for the bounce solutions and evaluating the corresponding Euclidean action, we further confirm this behavior in thick-wall cases. Our results indicate that overdensities can, in general, trigger vacuum decay earlier.}

\begin{document}
\maketitle
\flushbottom

\section{Introduction}\label{sec:introduction}

Vacuum decay plays a central role in particle cosmology, governing the occurrence of first-order phase transitions (FOPTs)~\cite{Mazumdar:2018dfl,Hindmarsh:2020hop,Caldwell:2022qsj} in the early Universe, which serves as an important source for stochastic gravitational-wave backgrounds (SGWBs) through bubble wall collisions~\cite{Hawking:1982ga,Witten:1984rs,Jinno:2016vai}, sound waves~\cite{Hogan:1986dsh,Hindmarsh:2016lnk,Hindmarsh:2019phv,Cai:2023guc}, and magnetohydrodynamic turbulences~\cite{Witten:1984rs,Kamionkowski:1993fg}. The impact of gravitational effects on FOPTs has been investigated to some extent in recent years, including the cosmic expansion effects~\cite{Cai:2018teh,Guo:2020grp,Jinno:2024nwb,Giombi:2025tkv} and the nonlinear interaction with primordial back holes (PBHs)~\cite{Jinno:2023vnr,Yuwen:2024gcf,Wang:2025hwc}.

Similar stories should apply not only to bubble dynamics but also to nucleations. While vacuum decay in flat spacetime has been extensively studied (see Ref.~\cite{Devoto:2022qen} for a detailed review), realistic cosmological settings inevitably involve gravitational effects. The presence of gravity not only modifies the structure of the tunneling solution but can also qualitatively change the decay rate, which was first demonstrated by Coleman and De Luccia~\cite{Coleman:1980aw} in the early 1980s, followed by the study on the effects from global spatial curvature carried by Hawking and Moss~\cite{Hawking:1981fz}. Subsequently, people began to consider the strong gravitational effects, including the catalytic effect from PBHs on vacuum decay processes~\cite{Moss:1984zf,Hiscock:1987hn} (see also some recent studies~\cite{Oshita:2019jan,Salvio:2016mvj,Vicentini:2022pra,Antoniadis:2024ent}). However, due to the emergence of the cone singularity during Wick rotation, mathematical tricks are required to evaluate its contribution to the action correctly~\cite{Gregory:2013hja,Burda:2015isa}. The results show that the decay rates can get enhanced for suitable parameter ranges~\cite{Burda:2015yfa}, which might be explained as the excitations due to Hawking radiation from the PBHs~\cite{Mukaida:2017bgd}. 
These developments highlight the importance of understanding vacuum decay beyond the flat-space approximation; meanwhile, they also naturally motivate the question of whether less extreme but more general gravitational effects, such as primordial curvature perturbations, can influence vacuum decay in the early Universe.

In this paper, we carry out a preliminary investigation of vacuum decay affected by curvature perturbations. For simplicity, we focus on a spatially spherically symmetric profile with $O(3)$ symmetry of the curvature perturbation, which captures the essential physics while remaining analytically and numerically tractable. The physical motivation for this assumption is stated as follows.
In the zero-temperature and high-temperature limits, the bounce solution has been shown to possess an $O(4)$ and $O(3)$ symmetry, respectively~\cite{Coleman:1977th,Linde:1980tt}, allowing the solution to be obtained by an effective one-dimensional ordinary differential equation that can be solved using the shooting method. There have been many public packages to search for the bounce solution with $O(N)$ symmetry for single-field or multi-field transitions~\cite{Masoumi:2017trx,Guada:2020xnz,Wainwright:2011kj}. 
Recently, it was mathematically proved that the bounce solution at finite temperature should still have an $O(3)$ symmetry~\cite{Shoji:2025nvj}, hence strengthening the assumption that it is sufficient to treat the system to be $O(3)$ symmetric~\cite{Masoumi:2012yy,Lee:2021nwg} and focus on the spherically symmetric case.

This paper is structured as follows. In Sec.~\ref{sec: Euclidean bounce}, we formulate the Euclidean bounce problem in the presence of curvature perturbations and discuss the dynamics of the corresponding bounce solutions under the zero-temperature and high-temperature limits. We then focus, in Sec.~\ref{sec: thin-wall}, on the properties of the bounce solution within the thin-wall approximation at the high temperature limit. In Sec.~\ref{sec: numerical res}, we present the numerical results of the bounce solutions at finite temperature. Finally, our conclusions and discussions are summarized in Sec.~\ref{sec:conclusions}. Throughout this paper, we take the natural units $\hbar = c = 1$.

\section{Euclidean bounce under curvature perturbation}\label{sec: Euclidean bounce}

In this section, we construct the Euclidean bounce solution in the presence of curvature perturbation. Here, we consider a system made up of gravity and a real scalar field serving as the phase transition field, 
\begin{align}
    S = \int \md^4 x\sqrt{-g} \left( \frac{R}{16\pi G} -\frac{1}{2}g^{\mu\nu}\nabla_\mu\phi \nabla_\nu\phi - V(\phi) \right)~.
\end{align}
For simplicity, the system is assumed to have spherical symmetry, with the vacuum bubble nucleated at the center of the spherical coordinate system.

\subsection{Lorentz-type spacetime setup}

For the gravity side, it was shown that on the super-horizon scale, the metric with curvature perturbation $\zeta(r)$ can be approximately written in an inhomogeneous but spatial isotropic coordinate~\cite{Shibata:1999zs},
\begin{align}\label{eq: initial metric}
    \md s^2 = -\md t^2 + a(t)^2 \me^{2\zeta(r)} \left( \md r^2 + r ^2 \md\Omega^2 \right)~. 
\end{align}
From numerical simulations~\cite{Joana:2025gqf}, one can see that the curvature perturbations grow up right after entering the horizon, followed by non-linear gravitational processes that might end up with the formation of PBHs or outgoing sound waves~\cite{Zeng:2025law,Ning:2025ogq,Ning:2025yvj}. Let us assume that the metric retains the form of~\eqref{eq: initial metric} at the time of the phase transition we are interested in. The spatial profile $\zeta(r)$ of curvature perturbation usually depends on specific physical processes and can be estimated by the two-point correlation function~\cite{Bardeen:1985tr}. For example, under the assumption of a narrow-peak scalar perturbation power spectrum on small scales $\mathcal{P}_\zeta(k) \simeq A_\zeta \delta(k - k_*)$, the mean profile is given by a sinc function~\cite{Yoo:2018kvb,Yoo:2020dkz},
\begin{align}\label{eq: sinc}
    \zeta(r) = \mu ~\mathrm{sinc}(k_*r)\equiv \mu \frac{\sin k_* r}{k_* r}~.
\end{align}
More generally, the profile may also be simply approximated as a Gaussian function for a given length scale $k_*^{-1}$~\cite{Musco:2018rwt},
\begin{align}\label{eq: exp}
    \zeta(r) = \mu \exp\left(-(k_* r)^2\right)~.
\end{align}
In this work, the discussions hold for any choice of $\zeta(r)$ as long as $\zeta$ decays to $0$ at $r\to\infty$, and in numerical calculations, we will compare the results with the forms~\eqref{eq: sinc} and~\eqref{eq: exp} as typical examples.

The FOPT can be illustrated by a real scalar field $\phi$ minimally coupled to gravity with the following typical polynomial potential~\cite{Lewicki:2019gmv},
\begin{align}\label{eq: potential}
    V(\phi) = V_T + (V_F -V_T)\left( 1 + \lambda \left(\frac{\phi}{\phi_T}\right)^2 - (2\lambda+4)\left(\frac{\phi}{\phi_T}\right)^3 + (\lambda+3)\left(\frac{\phi}{\phi_T}\right)^4 \right)~,
\end{align}
which has a true vacuum (global minimum) $V_T$ at $\phi=\phi_T$ and a false vacuum (local minimum) $V_F$ at $\phi=0$ with $V_T<V_F$, separated by a potential barrier described by a dimensionless parameter $\lambda$. Note that the potential~\eqref{eq: potential} is just a phenomenological parametrization, and the subsequent analysis and conclusions apply generally to any potential that allows an FOPT. If the initial spatial profile of the scalar field is provided, the classical time evolution is then controlled by the equation of motion (EOM)
\begin{align}\label{eq: EOM full}
    -\partial_t^2\phi - 3H\partial_t\phi + \frac{\me^{-2\zeta}}{a^2} \left( \partial_r^2\phi + \left(\frac{2}{r} + \zeta'\right) \partial_r\phi  \right) - \frac{\md V}{\md \phi} = 0~,
\end{align}
where $H=\md\ln a/\md t$ is the Hubble parameter, and the prime denotes the derivative with respect to $r$.

Here we are interested in sub-horizon vacuum bubbles. After the perturbations enter the horizon, the perturbations grow slowly with a timescale longer than several Hubble time, as is confirmed through numerical simulations~\cite{Joana:2025gqf}. By contrast, the FOPTs of interest should occur within a timescale much less than the Hubble time~\cite{Guth:1982pn,Turner:1992tz}, and the second term in Eq.~\eqref{eq: EOM full} describing friction from cosmic expansion can be safely neglected. For example, let us consider a typical parameter choice $\gamma / H \simeq \mathcal{O}(100)$, where $\gamma$ denotes the coefficient of time $t$ in the exponential factor of the vacuum decay rate, $\Gamma=\Gamma_0 \exp(\gamma t)$. The FOPT will roughly complete within $\gamma^{-1}\ll H^{-1}$, and thereby the background metric evolves negligibly during the phase transition~\footnote{
Let us take a rough estimation here. The linear perturbation equation of energy density contrast reads $\ddot{\delta} + 2H \dot{\delta} - \left( c_s^2 a^{-2} \nabla^2 + 4\pi G \rho \right) \delta = 0$.
The spatial gradient is of order of the Hubble scale $\nabla^2/a^2 \to -k^2/a^2 \sim -H^2$, and the gravitational source term $4\pi G \rho \sim 3H^2/2$. Then the equation can be recast to $\ddot{\delta} + 2H \dot{\delta} + (c_s^2- 3/2)H^2 \delta \simeq 0$. The characteristic time scale of this growth equation is of order $H^{-1}$. For large over-density region where the perturbation theory fails, the cosmic expansion term and pressure term are negligible compared to the gravity source, and the density contrast roughly follows the equation $\ddot{\delta}\sim 4\pi G\delta \rho \sim 3H^2\delta/2$, associated with a Jeans free-fall time $t\sim 1/\sqrt{4\pi G\delta \rho}$, which again implies a characteristic growth timescale of order $H^{-1}$. 
In both cases, it takes $\sim \mathcal{O}(H^{-1})$ for perturbation $\zeta$ to change significantly, while it takes $\sim \mathcal{O}(\gamma^{-1})\ll H^{-1}$ for an FOPT to complete.
}. 
In other words, the FOPT occurs much faster than the background evolution, allowing us to adopt the adiabatic approximation : the perturbation profiles can be treated as nearly static, with vacuum bubbles nucleating on top of such a non-flat spatial slice. 
We further assume that the spatial profile of $\zeta$ approximately preserves the shape it had on super-horizon scales, but with an enhanced amplitude due to the gravitational collapsing and a stretched wavelength due to the propagation of sound waves. In this way, we expect the evolution time scale of the perturbation to be much longer than the time scale for completing an FOPT, so that the metric can be approximately regarded as static, with its spatial inhomogeneity encoded by the curvature perturbation $\zeta$.
The scale factor is a dimensionless measure of the length scale and can be set to be unity here. Under these assumptions, the metric ansatz is then simplified into
\begin{align}\label{eq: metric}
    \md s^2 \simeq -\md t^2 + \me^{2\zeta(r)} \left( \md r^2 + r ^2 \md\Omega^2 \right),
\end{align}
and the EOM becomes
\begin{align}\label{eq: EOM}
    -\partial_t^2\phi + \me^{-2\zeta} \left( \partial_r^2\phi + \left(\frac{2}{r} + \zeta'\right) \partial_r\phi  \right) - \frac{\md V}{\md \phi} = 0~.
\end{align}

From a classical EOM~\eqref{eq: EOM}, if the scalar field was initially set up in the false vacuum $\phi=0$ with some small perturbations $\delta \phi/\phi_T \ll 1$, the kinetic energy is insufficient to support it to climb over the potential barrier and roll into the true vacuum~\footnote{In principle, even for the small perturbation but spanned over large enough region, it is still possible to eventually climb over the potential barrier by energy transfer from the gradient term to kinetic term via semi-classical channel~\cite{Wang:2019hjx}, though it usually takes more time to realize such transition, hence we will not consider this special case.}. If the initial profile interpolates smoothly between the false and true vacua, the pressure induced by the potential energy difference will drive the expansion of the true-vacuum region, converting vacuum energy into the kinetic and gradient energies of the scalar field. Classically, it is impossible to transfer the first case into the second case without any external interruptions, but this can happen from a quantum point of view.

\subsection{Instanton from Euclidean bounce}

The quantum nature of the scalar field suggests a tunneling from the false vacuum to the global minimum of the potential, resulting in a true vacuum bubble. Conventionally, the bubble nucleation process is associated with a decay rate $\Gamma\propto \exp(-S_\mathrm{E})$, where $S_\mathrm{E}=-iS$ is the Euclidean action of the system after performing Wick rotation $\tau=it$ with the following metric,
\begin{align}\label{eq: metric Euclidean}
    \md s_\mathrm{E}^2 = \md\tau^2 + \me^{2\zeta(r)} \left( \md r^2 + r ^2 \md\Omega^2 \right)~,
\end{align}
where the Euclidean time direction $\tau$ is compactified by a period equal to the inverse temperature $\beta=1/T$. 
The Euclidean action can be computed using the instanton approach~\cite{Coleman:1977py,Callan:1977pt,Linde:1981zj}. It should be remembered that a Wick rotation is not always applicable to any spacetime. If the background metric evolves with time (or to say does not have a global time-like killing vector), the system should be regarded as out of thermal equilibrium, in which case a real-time Schwinger–Keldysh closed-time path (CTP) integral, known as the in-in formalism, works more properly~\cite{Calzetta:2008iqa}. In our case, since we are working with the metric~\eqref{eq: metric} by neglecting cosmic expansion, the system is (near-)equilibrium, and the Euclidean path integral remains valid.

In this work, we focus on the dynamics of $\phi$ on a fixed background \eqref{eq: initial metric} by turning off the scalar-field back-reaction to gravity. Therefore, the Euclidean action that contributes to the decay rate only comes from the matter part. Taking into account the spherical symmetry of the system, the Euclidean action is a functional of the $\phi$ field given as
\begin{align}\label{eq: SE}
    S_\mathrm{E} = 4\pi \int \md \tau\md r~r^2\me^{3\zeta} \left( \frac{1}{2}(\partial_\tau \phi)^2 + \frac{1}{2} \me^{-2\zeta}(\partial_r \phi)^2 + V(\phi)\right)~.
\end{align}
Since the Lagrangian is invariant under time translation $\tau\to\tau + \tau_0$ and inversion $\tau\to -\tau$, without loss of generality, we can focus on the field in the region $0\leq\tau\leq\beta/2$ with boundary condition $\partial_\tau \phi|_{\tau=0}=0$. The Euclidean EOM is then given by
\begin{align}\label{eq: EOM Eulicdean}
    \partial_\tau^2\phi + \me^{-2\zeta} \left( \partial_r^2\phi + \left(\frac{2}{r} + \zeta'\right) \partial_r\phi  \right) - \frac{\md V}{\md \phi} = 0~.
\end{align}
An instanton is characterized by a ``bounce'' solution minimizing the Euclidean action \eqref{eq: SE} with the following boundary conditions (BCs),
\begin{align}\label{eq: BCs}
\begin{aligned}
    \mathrm{Outer~BC:} &\quad \phi(\tau,r\to\infty) = 0, \quad \phi(-\beta/2,r) = \phi(\beta/2,r)~, \\
    \mathrm{Inner~BC:} &\quad \phi(0,0) = \phi_0>0, \quad \partial_\tau \phi(0,r)=\partial_r \phi(\tau,0)=0~.
\end{aligned}
\end{align}
Before continuing to solve the EOM numerically, we discuss the two temperature limits. 

\subsubsection{High-temperature limit}\label{subsec: highT}

For high temperature limit, the period of imaginary time $\beta=1/T$ is required to be much smaller than the bubble spatial radius $R_c$, such that the bounce solution can safely reduce to an $O(3)$ solution. As discussed in the last section, we are interested in the vacuum bubbles whose size is much smaller than the Hubble radius, $R_c\ll H^{-1}$. Combining the two requirements above and using the Friedman Equation leads to
\begin{align}
    \frac{1}{T} = \beta \ll R_c \ll H^{-1} \sim \frac{M_p}{T^2} \quad \Rightarrow \quad T\ll M_p \sim 10^{18}\,\mathrm{GeV}~,
\end{align}
where $M_p$ is the reduced Planck mass. Therefore, a sub-horizon bubble can be realized provided that the temperature is below the Planck scale. This condition is not difficult to satisfy, for example, some BSM models at $T\sim 100\,\mathrm{GeV}$ for electro-weak FOPTs

Under this limit, the $S^1$ topology of the Euclidean time direction degenerates to a trivial point, so that one can safely drop the $\tau$ dependence of $\phi$, yielding an ordinary differential equation that has an $O(3)$ symmetry rather than a partial one,
\begin{align}
    \phi'' + \left(\frac{2}{r} + \zeta'\right)\phi' - \me^{2\zeta}\frac{\md V}{\md \phi} = 0~.
\end{align}
Solving the EOM above through the shooting method to satisfy the boundary conditions $\phi'(0)=0$ and $\phi(\infty)=0$, one can obtain a bounce solution interpolating between the false vacuum $\phi=0$ and a turning point $\phi_0$, the latter of which lies on the true vacuum side of the potential barrier. 

\begin{figure}
    \centering
    \includegraphics[width=0.95\linewidth]{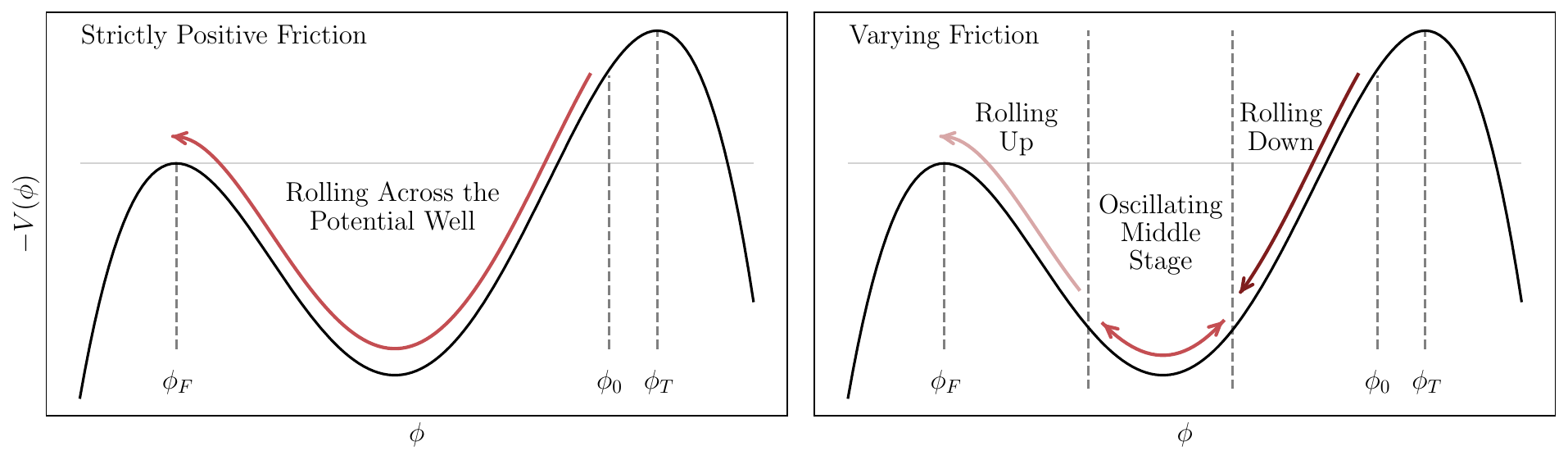}
    \caption{An illustrative picture of the motion of a scalar field in Euclidean spacetime, where the turning point is denoted as $\phi_0$. The left panel shows a simple rolling from $\phi_0$ to $\phi_F$, which corresponds to a ``bubble wall'' configuration. The right panel shows the existence of an oscillating middle stage by allowing the friction term to vary in sign.}
    \label{fig: potential}
\end{figure}

In the case $\zeta=0$, the turning point should have a lower potential energy $V(\phi_0)<V_F$, the reason of which is stated as follows: from the Euclidean spacetime point of view, the instanton corresponds to a particle rolling down from $\phi_0$ at rest under an effective potential $-V$, where the ``time'' parameter is $r$~\footnote{One may extend the solution to the negative half-axis $r<0$. In this case, the particle rolls down from the false vacuum as $r\to-\infty$, reaches the turning point at $r=0$, and then returns to the false vacuum as $r\to\infty$, hence the name ``bounce''.}. Then multiplying the EOM with $\phi'$ results in
\begin{align}\label{eq: cons. h-T flat}
    \frac{\md}{\md r}\left( \frac{1}{2}(\phi')^2 - V\right) = -\frac{2}{r}(\phi')^2~.
\end{align}
If there were no friction term $2(\phi')^2/r$ on the right-hand side, the summation of kinetic energy $(\phi')^2/2$ and potential $-V$ is conserved during the evolution along $r$. It is because of the existence of the strictly positive friction term that requires the particle’s initial energy $-V(\phi_0)$ greater than its final energy $-V_F$, i.e., $V(\phi_0)<V_F$. See the left panel of Fig.~\ref{fig: potential} for an illustrative picture.

\begin{figure}
    \centering
    \includegraphics[width=\linewidth]{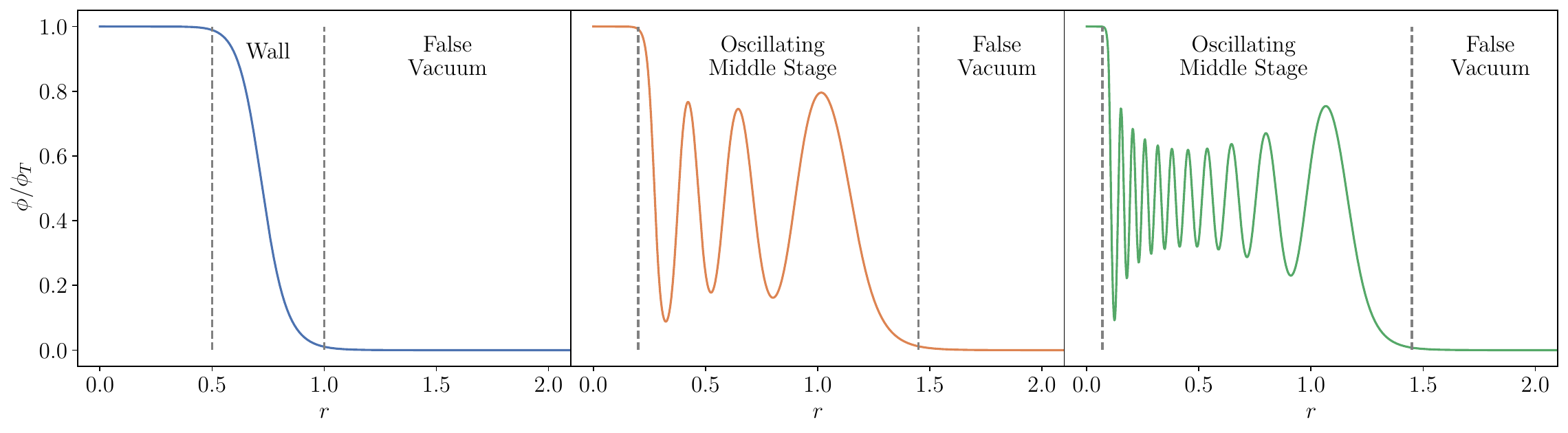}
    \caption{The bounce solution under curvature perturbation $\zeta=\mu \exp(-r^2)$ with $\mu=1$ (left), $2$ (middle), and $3$ (right), respectively, where the potential barrier parameter is set to be $\lambda=50$. In the left panel, the condition $r^{-1}+\zeta'$ is always satisfied, and hence there is a bubble wall connecting the two vacuums, while the right two panels exhibit an oscillating middle stage.}
    \label{fig: high-T}
\end{figure}

However, a general $\zeta$ no longer guarantees the conservation law \eqref{eq: cons. h-T flat} but leads to a similar one after re-parametrization. Let us define
\begin{align}\label{eq: rt def}
    \tilde{r} \equiv \int_0^r \md x \exp(\zeta(x))~,\quad ~ \frac{\md \tilde{r}}{\md r} = \exp(\zeta(r))~,
\end{align}
in which case, the EOM can be recast as
\begin{align}\label{eq: EOM phidot}
    \ddot{\phi} + 2\me^{-\zeta} \left(\frac{1}{r} + \zeta' \right)\dot{\phi} - \frac{\md V}{\md \phi} = 0~.
\end{align}
where the dot and prime stands for the derivative over $\tilde{r}$ and $r$, respectively. Multiplying the EOM with $\dot{\phi}$ yields a new ``conservation law'',
\begin{align}\label{eq: cons. h-T}
    \frac{\md}{\md \tilde{r}}\left( \frac{1}{2}\dot{\phi}^2 - V\right) = -2\me^{-\zeta}\left(\frac{1}{r} + \zeta'\right) \dot{\phi}^2 ~.
\end{align}
For the case $r^{-1}+\zeta'$ is always positive, one can reach the same conclusion $V(\phi_0)<V_F$ as before. This may break down for large positive curvature perturbations associated with a negative gradient $\zeta'$, for example $\zeta=\mu \exp(-r^2)$ with a large enough $\mu$. Although theoretically possible, a very large amplitude for curvature perturbation is not a common case and may lead to PBH formation. Therefore, in most cases, the condition $V(\phi_0)<V_F$ remains valid.

It is known to us that for a strictly positive $r^{-1}+\zeta'$, the profile takes the form of a ``bubble wall'' configuration, as is shown in the left panel of Fig.~\ref{fig: high-T}. However, the story changes when this friction term is allowed to change the sign:
Initially at $r\to 0$, the divergence behaviour of $r^{-1}$ ensures the condition $r^{-1}+\zeta'>0$, providing a friction that dissipates the particle's energy. Gradually, the negative $\zeta'$ begins to dominate over $1/r$ and leads to a driving force that is proportional to the velocity $\dot{\phi}$, pulling the particle oscillating around the trough of $-V$ (the peak of the potential barrier in $V$). After certain time scale, $\zeta'$ decays back to $0$ and $r^{-1}$ begin to dominate again. Such an oscillating middle stage is found numerically and shown in the right two panels in Fig.~\ref{fig: high-T}. At last, the particle asymptotically approaches the false vacuum, thereby satisfying the boundary conditions at $r\to\infty$. An intuitive picture is shown in the right panel of Fig.~\ref{fig: potential}.

\subsubsection{Zero-temperature limit}

For the zero-temperature limit $\beta\to \infty$, the periodic boundary condition along $\tau$ degenerates into a similar boundary condition to $r$ at infinity, $\phi(\tau\to\pm\infty,r) = 0$. In the simplest case with $\zeta=0$, the time direction reduces to nothing but another equivalent spatial direction, and the system recovers an $O(4)$ symmetry by introducing a global ``radial'' coordinate $\rho = \sqrt{\tau^2 + r^2}$, yielding a simplified EOM described by another ordinary-differential equation that again can be solve by the shooting method,
\begin{align}
    \frac{\md^2\phi}{\md\rho^2} + \frac{3}{\rho}\frac{\md\phi}{\md\rho} - \frac{\md V}{\md \phi} = 0~.
\end{align}
However, this in general cannot be achieved for an arbitrary $\zeta$, which can be shown by analyzing the geometric properties of \eqref{eq: metric Euclidean}. Let us assume there exists a scalar function $\rho(\tau,r)$ that rewrites the metric to have a $O(4)$ symmetry, $\md s^2 = F(\rho)\left(\md\rho^2+\rho^2\md\Omega_{(3)}^2 \right)$, where $F(\rho)$ is a non-negative function of $\rho$ and $\md\Omega_{(3)}^2$ is the standard line-element of a hyper-sphere $S^3$. Apparently, the aimed metric is conformally flat with a vanishing Weyl tensor, while the Weyl tensor of metric \eqref{eq: metric Euclidean} is proportional to a functional of $\zeta$, therefore it's equivalent to require
\begin{align}
    C_{\mu\nu\sigma\lambda} \propto \zeta''(r) - \zeta'(r)^2 - \frac{\zeta'(r)}{r} = 0~.
\end{align}
There are two independent solutions to the equation above,
\begin{align}
    \zeta_1(r) = C_1 - \ln(r^2+C_2),\quad \zeta_2(r)=\mathrm{const.},
\end{align}
where $C_1$ and $C_2$ are two integral constants. On the one hand, $\zeta_1$ diverges as $r\to\infty$, which is contradictory to our initial assumptions that $\zeta$ decays to $0$. On the other hand, a constant $\zeta_2$ just corresponds to a trivial rescaling of radial direction $r$, which brings us back to a case equivalent to $\zeta = 0$. Therefore, a physical choice of non-constant $\zeta$ always breaks the $O(4)$ symmetry at the zero-temperature limit.

\section{The thin-wall approximation at high temperature}\label{sec: thin-wall}

In this section, we conducted an analytical study of the thin-wall approximation, which arises when the potential barrier is much larger than the energy difference between the two vacuums. Let us denote $R$ as the radius of the vacuum bubble and $\delta$ as the width satisfying $\delta\ll R$. The physical picture of the thin-wall approximation can be divided into three stages:
\begin{enumerate}
    \item The scalar field starts to slowly roll down from an initial position close to the true vacuum $\phi_0\simeq\phi_T$. Because of the balance between a near-vanishing driving force $-\md V/\md\phi$ and a strong friction proportional to the velocity $\dot{\phi}$, the scalar field stays close to $\phi_T$ for a long time until $r\simeq R$.
    \item The friction can not stop the scalar field from rolling down any longer, and the scalar field undergoes a rapid roll through a deep potential well in $-V$ (a tall potential barrier in $V$) within a short period $\Delta r\simeq \delta$, corresponding to the thin-wall region in Lorentz spacetime.
    \item The kinetic energy transfers back to potential energy, and the scalar field finally stops at the false vacuum as $r\to\infty$.
\end{enumerate}
Later in this section, only the high-temperature limit is considered, where there is an $O(3)$ symmetry in the bounce solution. Different from the flat case, an $O(4)$ symmetry no longer holds in the zero-temperature limit, which introduces difficulties for an analytical study and therefore is not considered here.
In addition, we only focus on the case where $\zeta'$ is small, such that an oscillating middle stage described in Sec.~\ref{subsec: highT} is avoided, which is also a physically well-motivated assumption as curvature perturbations are typically small in most situations.

\subsection{Euclidean action}

To begin with, it should be noted that the actual value contributing to the vacuum decay rate is the difference between the Euclidean action of the bounce solution and that of the classical static solution $\phi=\phi_F \equiv 0$~\cite{Callan:1977pt,Linde:1980tt}. Without the subtraction of the static solution, the action will be divergent since the radial coordinate $r$ is unbounded. Consequently, in the high-temperature limit, the Euclidean action relevant for our analysis takes the form
\begin{align}
    \Delta S_\mathrm{E} := S_\mathrm{E}[\phi] - S_\mathrm{E}[\phi_F] = 4\pi \beta \int \md r~r^2\me^{3\zeta} \left( \frac{1}{2} \me^{-2\zeta}(\phi')^2 + V(\phi) - V_F \right)~,
\end{align}
which can be divided into three contributions: the interior of the bubble, the bubble wall, and the exterior. The last one gives a vanishing contribution because the false vacuum potential has been subtracted throughout the space. Then the action can be approximated as
\begin{align}\label{eq: DSE}
    \Delta S_\mathrm{E} \simeq 4\pi \beta\left(R^2\me^{2\zeta(R)} \sigma - \int_0^R\md r ~r^2\me^{3\zeta} \Delta V\right)~,
\end{align}
where $\Delta V = V_F - V_T$ is the potential difference and $\sigma$ is the tension of the bubble wall defined as
\begin{align}\label{eq: sigma def}
    \sigma = \int_\mathrm{wall} \md r ~\me^{\zeta(r)} \left( \frac{1}{2} \me^{-2\zeta}(\phi')^2 + V(\phi) - V_F\right) = \int_\mathrm{wall} \md \tilde{r} \left( \frac{1}{2}\dot{\phi}^2 + V(\phi) - V_F\right)~,
\end{align}
with $\tilde{r}$ defined in Eq.~\eqref{eq: rt def}. Our goal is to find the relation among $R$, $\Delta V$, and $\sigma$, and then find a general procedure to evaluate the action for a given potential.

To evaluate the tension $\sigma$, we refer back to the EOM~\eqref{eq: EOM phidot}. Since we're working on the thin-wall approximation, the friction term is negligible in the vicinity of the bubble wall, yielding a conservation
\begin{align}
    \frac{\md}{\md \tilde{r}}\left( \frac{1}{2}\dot{\phi}^2 - V\right) \simeq 0 \quad \mathrm{at} \quad R\leq r \leq R+\delta~.
\end{align}
By integrating the equation above and aligning the asymptotic behavior $\dot{\phi}\to0$ and $V\to V_F$ at $r\to\infty$, one obtains a first integral corresponding to an energy-like conserved quantity $-V_F$,
\begin{align}
    \frac{1}{2}\dot{\phi}^2 - V = -V_F
\end{align}
Substituting back to the definition \eqref{eq: sigma def},
\begin{align}\label{eq: sigma}
    \sigma = \int_\mathrm{wall} \md \tilde{r} ~\dot{\phi}^2 = \int_\mathrm{wall} \md \phi~ \frac{\md\phi}{\md\tilde{r}} = \int_{\phi_F}^{\phi_t} \mathrm{d}\phi ~\sqrt{2(V-V_F)}~,
\end{align}
where $\phi_t$ is the asymptotic value of the scalar field toward the true vacuum side of the bubble wall, which is slightly smaller than $\phi_T$ to ensure the positive definiteness of $V-V_F$ in the square root. Under the thin-wall limit where $\Delta V$ is much smaller than the potential barrier, the integration in Eq.~\eqref{eq: sigma} roughly covers the interval from $\phi_F$ to $\phi_T$.
Therefore, the tension $\sigma$ is uniquely determined by the shape of the potential barrier, and is independent of the initial profile of the vacuum bubble and the spatial geometry of the background spacetime. This exactly reduces to the definition of $\sigma$ in the flat case~\cite{Coleman:1977py}. 

Since the bounce solution is an extremum of the action in the whole space of field configurations, the bubble wall radius $R$ in the thin-wall approximation can be determined by searching for the saddle point of \eqref{eq: DSE}. Requiring $\md\Delta S_\mathrm{E} / \md R = 0$ results in
\begin{align}\label{eq: Rflat}
    \frac{R\me^{\zeta(R)}}{1+ R\zeta'(R)} = \frac{2\sigma}{\Delta V} = R_\mathrm{flat}~,
\end{align}
where $R_\mathrm{flat}$ is defined as the bubble radius with $\zeta \equiv 0$. This result can be understood from another perspective. Integrating over the total conservation equation~\eqref{eq: cons. h-T} without neglecting the friction term, one can arrive at
\begin{align}
\begin{aligned}
    0 ~&= \int_0^\infty \md\tilde{r} \left[ \frac{\md}{\md \tilde{r}}\left( \frac{1}{2}\dot{\phi}^2 - V\right) + 2\me^{-\zeta}\left(\frac{1}{r} + \zeta'\right) \dot{\phi}^2 \right] \\
    &\simeq \left( \frac{1}{2}\dot{\phi}^2 - V\right)\bigg|_0^\infty + 2\me^{-\zeta(R)}\left(\frac{1}{R} + \zeta'(R) \right) \int_\mathrm{wall} \md\tilde{r}~\dot{\phi}^2 \\
    &= -\Delta V + 2\me^{-\zeta(R)}\left(\frac{1}{R} + \zeta'(R) \right)\sigma~,
\end{aligned}
\end{align}
where in the second line we have used the fact that $\dot{\phi}$ is non-zero only in the vicinity of the bubble wall, and in the last line we have used \eqref{eq: sigma}, as well as the asymptomatic behaviours of the bounce solution: $\dot{\phi}\to 0$ on the two boundaries, $V\to V_T$ and $V_F$ at $r\to 0$ and $\infty$ respectively. This is equivalent to the result~\eqref{eq: Rflat} we got from the variational principle.

Finally, the Euclidean action can be recast to
\begin{align}
    \Delta S_\mathrm{E} = 4\pi \left(y^2 - \frac{2}{y}(1+r\zeta') \mathcal{V}\right) \beta\sigma
\end{align}
with abbreviation $y=R\me^{\zeta(R)}$ and $\mathcal{V} = \int_0^R r^2\exp(3\zeta)\md r$ as the physical volume of the vacuum bubble.

\subsection{Comparison with the flat geometry}\label{subsec: thin-wall compare}

In the flat geometry case $\zeta=0$, the physical volume of the bubble can be analytically calculated, and further simplified for the Euclidean action is given by
\begin{align}\label{eq: SE flat}
    \Delta S_{E}^\mathrm{flat} = 4\pi R^2_\mathrm{flat} \beta\sigma - \frac{4\pi}{3} R^3_\mathrm{flat}\beta\Delta V = \frac{4\pi}{3}  R^2_\mathrm{flat} \beta\sigma~,
\end{align}
where the physical meaning of each term is clearer: the first term corresponds to the surface energy associated with the wall tension $\sigma$, and the second term is the vacuum energy within the bubble. 

In this subsection, the curvature perturbation takes the ansatz $\zeta = \mu f(k_* r)$, which is a monotonic function of $r$ satisfying $f(0)=1$ and $f(\infty)=0$~\footnote{The boundary values of $f$ are chosen such that the amplitude of curvature perturbation can be measured simply by $\mu$. For example, one can take $f(x)=\exp(-x^2)$, which recovers our ansatz~\eqref{eq: exp}.}, where $1/k_*$ is a characteristic length scale of the curvature perturbation. Then we compare the corresponding Euclidean action with the flat case by considering two limiting cases. The conclusions hold for any monotonic profile of $\zeta$. If $k_*R\ll 1$, the vacuum bubble is nucleated deep inside the curvature perturbation where $\zeta\simeq\mu$ and $\zeta'\simeq 0$. The relation \eqref{eq: Rflat} can be rewritten as
\begin{align} \label{eq: y and R}
    y = R\me^{\mu} \simeq R_\mathrm{flat} ~.
\end{align}
In this case, the physical volume of the bubble can be  integrated analytically,
\begin{align}
    \mathcal{V} = \int_0^R \left(\me^\mu r\right)^2 \me^\mu\md r \simeq \frac{1}{3}y^3~.
\end{align}
Then the Euclidean action can be approximated as
\begin{align}
\begin{aligned}
    \Delta S_\mathrm{E}^{k_*R\ll 1} = 4\pi \left(y^2 - \frac{2}{R_\mathrm{flat}} \mathcal{V}\right) \beta\sigma \simeq \frac{4\pi}{3}R_\mathrm{flat}^2 \beta\sigma~.
\end{aligned}
\end{align}
Together with \eqref{eq: SE flat} and \eqref{eq: y and R}, we can conclude that $\Delta S_\mathrm{E}^{k_*R\ll 1} \simeq \Delta S_{E}^\mathrm{flat}$.

On the other hand, if $k_* R \gg 1$, the radius of the bubble wall is much larger than the characteristic scale of the curvature perturbation. In this case, we have $\zeta\simeq 0$ and $\zeta'\simeq 0$ in the vicinity of the bubble wall, suggesting that $R\simeq R_\mathrm{flat}$. The non-trivial profile of $\zeta$ only affects the vacuum energy within the bubble by deviating its physical volume from the flat case,
\begin{align}
    \mathcal{V} = \int_0^R r^2\me^{3\zeta}~\md r \left\{ \begin{array}{cc}
        >R^3/3~, & ~\mu>0 \\
        <R^3/3~, & ~\mu<0
    \end{array} \right. \quad.
\end{align}
Then the Euclidean action is given by
\begin{align}
    \Delta S_\mathrm{E}^{k_*R\gg 1} \simeq 4\pi \left(R^2 - \frac{2}{R_\mathrm{flat}} \mathcal{V}\right) \beta\sigma \left\{ \begin{array}{cc}
        <\Delta S_\mathrm{E}^\mathrm{flat}~, & ~\mu>0 \\
        >\Delta S_\mathrm{E}^\mathrm{flat}~, & ~\mu<0
    \end{array} \right. \quad.
\end{align}
Therefore, in both limits we obtain a smaller Euclidean action for an over-density and a larger one for an under-density. As two conclusions are derived in the two extreme limits, it is natural to expect an intermediate behaviour to apply to the case in between, i.e. $\Delta S_\mathrm{E}^{k_*R\gg 1} < \Delta S_\mathrm{E}^{k_*R\simeq 1} < \Delta S_\mathrm{E}^{k_*R\ll 1} \simeq \Delta S_\mathrm{E}^\mathrm{flat}$. Without further rigorous mathematical proof, we believe that in general an over-density always results in a smaller Euclidean action and thus a higher vacuum decay rate in the thin-wall limit.

\section{Numerical results at finite temperature}\label{sec: numerical res}

In this section, we solve the Euclidean EOM \eqref{eq: EOM Eulicdean} numerically at finite temperature. Without an equivalence between the imaginary time direction and radial direction, it's impossible to reduce the EOM to an effective 1D equation, and the system must be solved directly on a 2D grid with boundary conditions \eqref{eq: BCs}. 
Since the equation is a nonlinear elliptic partial differential equation, it can be solved using Newton's method with a suitable initial guess for the solution.

\begin{figure}[htbp]
    \centering
    \includegraphics[width=\linewidth]{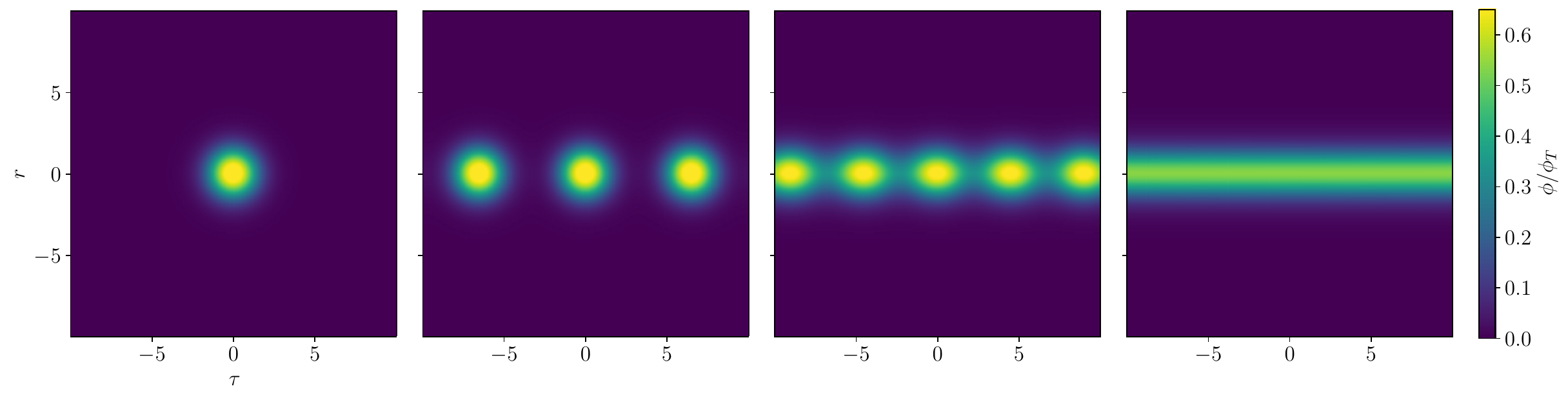}
    \includegraphics[width=\linewidth]{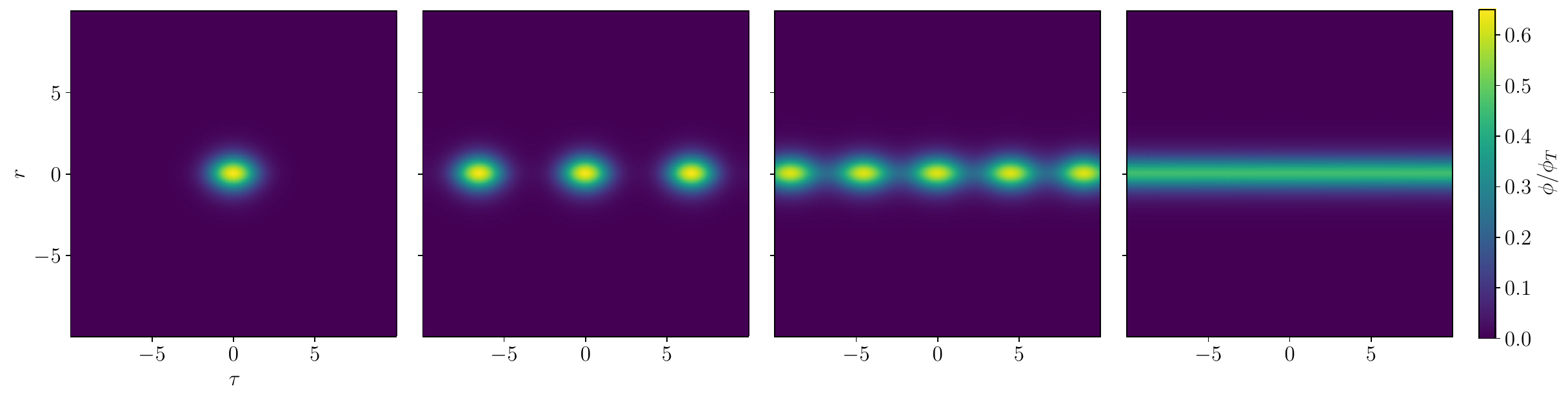}
    \includegraphics[width=\linewidth]{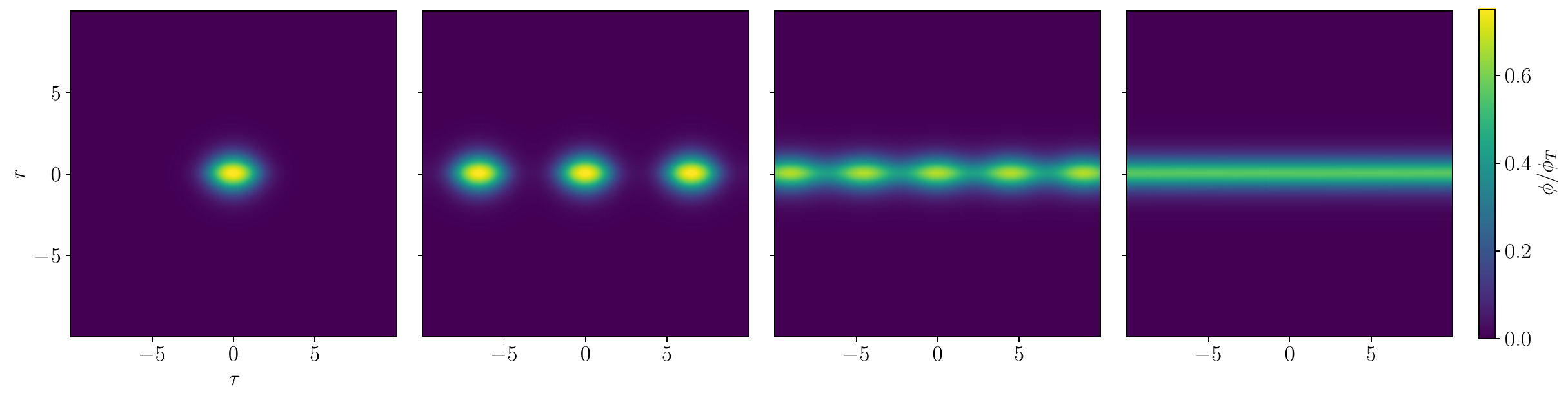}
    \includegraphics[width=\linewidth]{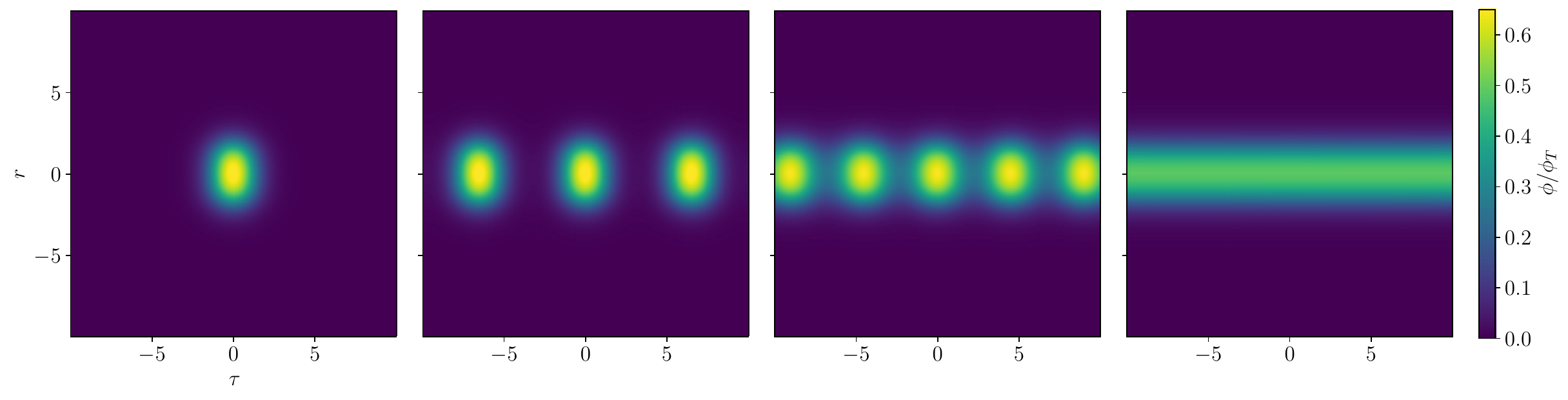}
    \includegraphics[width=\linewidth]{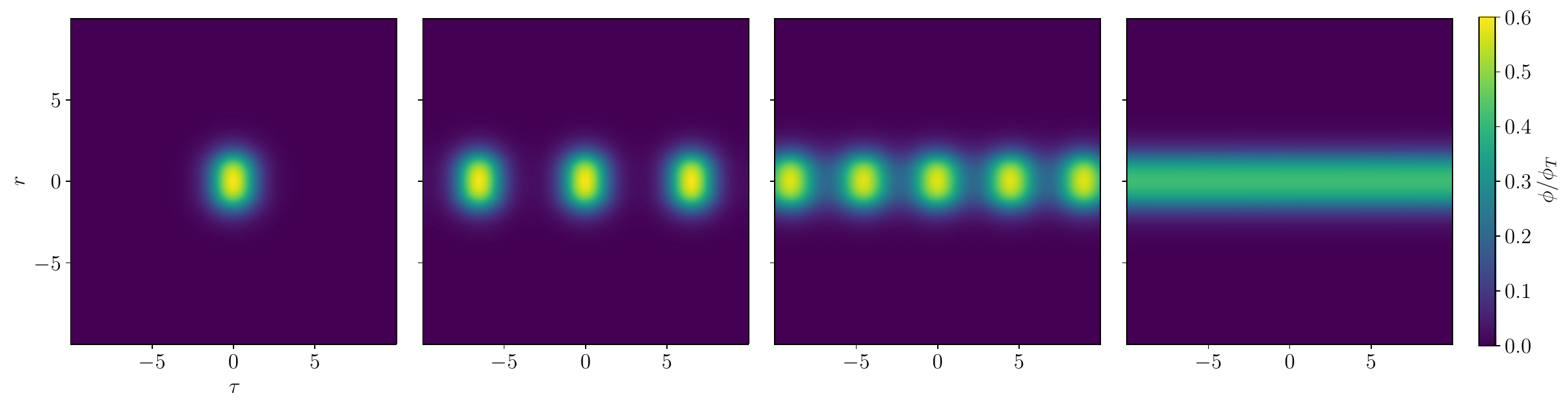}
    \caption{Numerical results of the bounce solutions under different curvature perturbation profiles with $\zeta=0$, $\mu\exp(-r^2)$, $\mu\mathrm{sinc}(\pi r)$, $-\mu\exp(-r^2)$, and $-\mu\mathrm{sinc}(\pi r)$, for $\mu=1/2$, from top to bottom, respectively. Each column corresponds to a different temperature with $\beta=\infty$, $6.5$, $4.5$, and $0$, from left to right, respectively.}
    \label{fig: 2D solution}
\end{figure}
\begin{figure}
    \centering
    \includegraphics[width=0.99\linewidth]{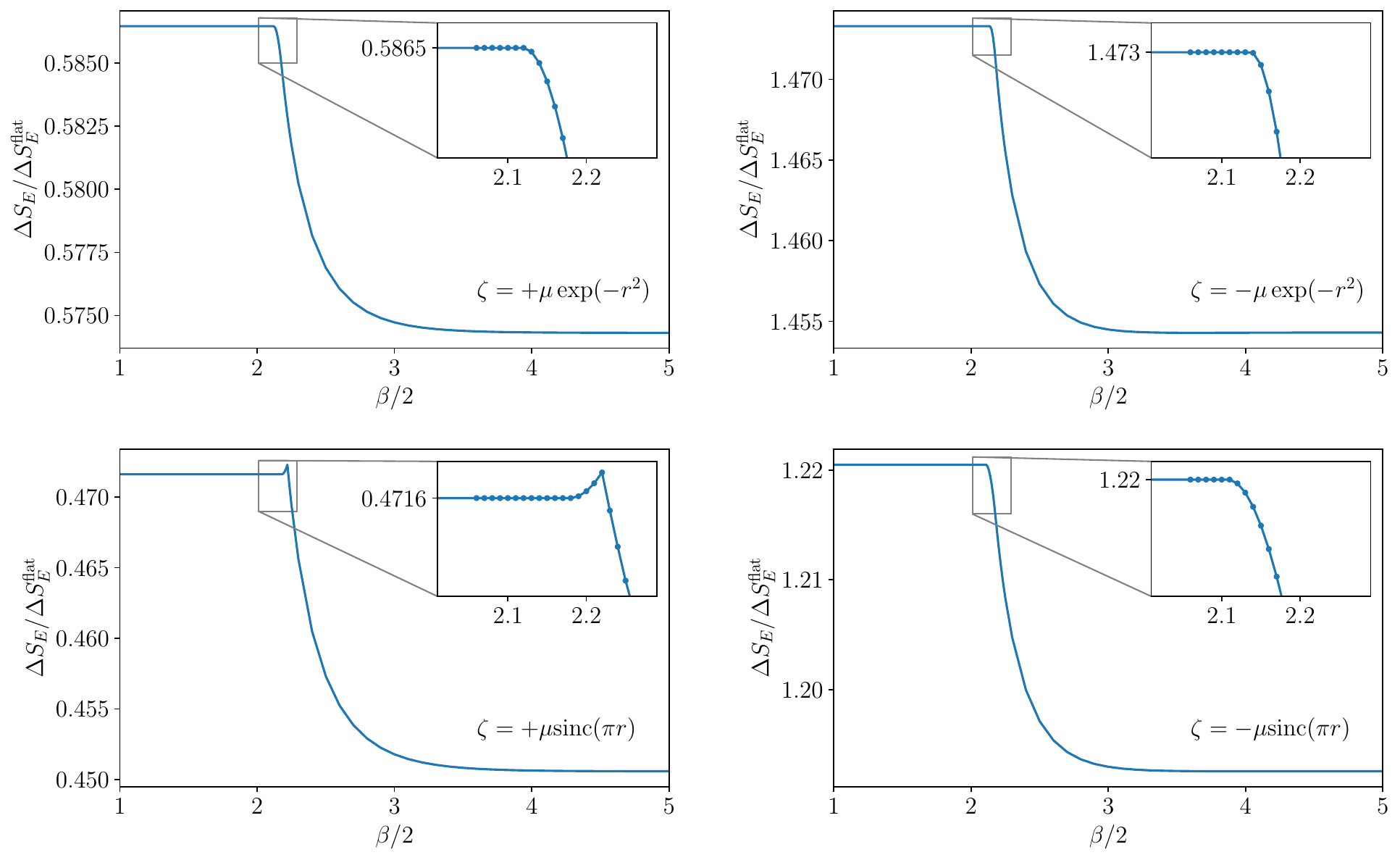}
    \caption{the ratio of the Euclidean action with curvature perturbations to that in the flat case as a function of inverse temperature $\beta=1/T$, with $\mu=1/2$ corresponding to the examples shown in Fig.~\ref{fig: 2D solution}.}
    \label{fig: action}
\end{figure}

The numerical implementation is stated as follows. First of all, the characteristic length scale of the system $1/k_*$ is to be unity. For generality, the potential parameters are set to be $\Delta V = 1$ and $\lambda = 1$, corresponding to a thick bubble wall configuration. Then, a 2D meshgrid is generated over the rectangular region $0\leq \tau\leq \beta/2$ and $0\leq r \leq 10$, with reflective boundary conditions along all edges. Although such a boundary condition is not exactly equivalent to \eqref{eq: BCs} at the outer boundary of $r$, it is satisfied to high accuracy and greatly simplifies our numerical implementation. For some special cases where the $O(3)$ or $O(4)$ symmetry is restored, we compare our results against those from publicly available code \texttt{AnyBubble}~\cite{Masoumi:2017trx} to validate our calculations. We start from solving the $O(4)$ solution for the flat case, which serves as the initial guess for the solution $\phi_{\beta_M}$ with a generic $\zeta$, where $\beta_M$ is chosen to be sufficiently large so that the solution effectively corresponds to the zero-temperature limit. Then we construct an inverse-temperature ladder $\beta_M, \beta_M-\Delta\beta,\cdots, \beta_m$ to label the solutions at different temperatures, reaching a relatively small $\beta_m$. For a given period $\beta_i$, the solution $\phi_{\beta_i}$ is obtained by performing a Newton iteration on the 2D grid, which again is set to be the initial guess for the solution with period $\beta_i - \Delta \beta$.

We consider the following cases: $\zeta=0$, $\mu \exp(-r^2)$ and $\mu~\mathrm{sinc}(\pi r)$, with both positive and negative perturbation $\mu=\pm 1/2$. The numerical solutions are shown in Fig.~\ref{fig: 2D solution}. For a more intuitive visualization, we extend $r$ to the negative half-axis, such that the entire $r$-axis corresponds to an arbitrary spatial direction passing through the coordinate origin. As we can see, the top-left corner shows the solution with $O(4)$ symmetry for the flat case, where the solution shows exactly the same dependence on $\tau$ and $r$. In the last column, we show the $O(3)$ solution at the high-temperature limit, where the scalar field becomes entirely time-independent. There are two main findings concluded from the numerical results:

\begin{enumerate}
    \item Although different profiles lead to slight quantitative differences in the numerical solution, the positive curvature perturbation ($\mu>0$) shown in the second and the third rows always causes the solution to be squeezed along the $r$ direction, corresponding to a smaller initial bubble radius. This behavior holds across different temperatures. Conversely, a negative perturbation, illustrated in the bottom two rows, always stretches the solution along $r$, resulting in a larger initial radius. This conclusion is consistent with our discussion on $R$ under the thin-wall approximation obtained in Sec.~\ref{subsec: thin-wall compare}. The ratios of the Euclidean actions with curvature perturbations to those in the flat case are shown in Fig.~\ref{fig: action}, where the actions are roughly halved $\Delta S_\mathrm{E}\to\Delta S_\mathrm{E}/2$ for over-densities. By noticing that in the beginning of the FOPT the decay rate $\Gamma$ should be much smaller than $1$ and the relation $\Gamma\sim\exp(-\Delta S_\mathrm{E})$, the results suggest that the decay rate is approximately enhanced by a square-root $\Gamma\to\Gamma^{1/2}$.
    \item Intuitively thinking, one may expect a smooth interpolation between the solution in the high-temperature limit, which has an $O(3)$ symmetry, and the one in the zero-temperature limit (recovering an $O(4)$ symmetry in the flat case). However, in numerical solutions, we find that, for each specific choice of curvature perturbation profiles, there exists a certain critical temperature $T_0$, above which the solution degenerates exactly back to the $O(3)$ solution. This critical temperature varies slightly for different curvature profiles. As is shown in Fig.~\ref{fig: action}, the ratio for each case saturates at a non-zero value of $\beta_0$ that corresponds to $T_0$.
\end{enumerate}

\section{Conclusions}\label{sec:conclusions}

In this work, we present a systematic study of vacuum decays in the presence of curvature perturbations $\zeta$ by constructing the corresponding bounce solutions and evaluating the corresponding Euclidean action. We first discuss how the EOM of the scalar field is modified by gravity, and study the symmetries and behaviors of the solutions under two limits. In the high-temperature limit, the solution gains an $O(3)$ symmetry as in the flat case. A middle stage oscillating at the peak of the potential barrier is found to gradually emerge in the vicinity of the bubble wall when $r^{-1}+\zeta'<0$ is satisfied for some range of $r$. In the zero-temperature limit, the presence of gravity prevents the solution from restoring an $O(4)$ symmetry by breaking the equivalence between the Euclidean time $\tau$ and the radial coordinate $r$.

In particular, we focus on the thin-wall approximation in the high-temperature limit, in which the Euclidean action admits a clear separation into contributions from the wall surface tension and the bulk vacuum energy inside the bubble. Requiring the variational principle (or equivalently the Euclidean EOM), the relation between bubble wall radius $R$, the tension $\sigma$, and the vacuum energy difference $\Delta V$, the last two of which can be determined uniquely by the shape of the potential $V$. We show that, in both the large- and small-bubble limits, an over-density always results in a smaller bubble radius $R$ and smaller Euclidean action $\Delta S_\mathrm{E}$, whereas an under-density results in a larger $R$ and $\Delta S_\mathrm{E}$, indicating that an over-density can trigger an earlier occurrence of an FOPT. 

At last, we numerically solve the EOM to find the Euclidean bounce solutions with a thick wall at finite temperature, and again obtain similar conclusions as in the high-temperature limit, which further supports the idea that FOPTs can, in general, be triggered by overdensities. A critical temperature is found in our numerical results, above which the solution degenerates to the high-temperature limit regardless of the value of $\beta$. An analytical study on this critical behavior is left as future work. In addition, since our numerical analysis is restricted to two specific spatial configurations of curvature perturbations, future studies can investigate how different perturbation profiles affect the decay rates and initial bubble profiles, and consequently influence the resulting SGWB signals.

\acknowledgments
R.-G. Cai and S.-J. Wang are supported by the National Key Research and Development Program of China Grants No. 2021YFC2203004, No. 2021YFA0718304, and No. 2020YFC2201501, the National Natural Science Foundation of China Grants No. 12422502, No. 12547110, No.12588101, No. 12235019, and No. 12447101, and the Science Research Grants from the China Manned Space Project with No. CMS-CSST-2025-A01.
Z.-Y. Yuwen is supported by an appointment to the Young Scientist Training (YST) program at the APCTP through the Science and Technology Promotion Fund and Lottery Fund of the Korean Government. This was also supported by the Korean Local Governments-Gyeongsangbuk-do Province and Pohang City.


\bibliographystyle{JHEP}
\bibliography{ref}

@article{Zeng:2025law,
    author = "Zeng, Xiang-Xi and Ning, Zhuan and Yuwen, Zi-Yan and Wang, Shao-Jiang and Deng, Heling and Cai, Rong-Gen",
    title = "{Relic gravitational waves from primordial gravitational collapses}",
    eprint = "2504.11275",
    archivePrefix = "arXiv",
    primaryClass = "gr-qc",
    month = "4",
    year = "2025"
}

@article{Ning:2025ogq,
    author = "Ning, Zhuan and Zeng, Xiang-Xi and Yuwen, Zi-Yan and Wang, Shao-Jiang and Deng, Heling and Cai, Rong-Gen",
    title = "{Sound waves from primordial black hole formations}",
    eprint = "2504.12243",
    archivePrefix = "arXiv",
    primaryClass = "gr-qc",
    month = "4",
    year = "2025"
}

@article{Ning:2025yvj,
    author = "Ning, Zhuan and Yuwen, Zi-Yan and Zeng, Xiang-Xi and Cai, Rong-Gen and Wang, Shao-Jiang",
    title = "{Acoustic gravitational waves from primordial curvature perturbations}",
    eprint = "2512.21151",
    archivePrefix = "arXiv",
    primaryClass = "gr-qc",
    month = "12",
    year = "2025"
}

@article{Wang:2019hjx,
    author = "Wang, Shao-Jiang",
    title = "{Occurrence of semiclassical vacuum decay}",
    eprint = "1909.11196",
    archivePrefix = "arXiv",
    primaryClass = "gr-qc",
    doi = "10.1103/PhysRevD.100.096019",
    journal = "Phys. Rev. D",
    volume = "100",
    number = "9",
    pages = "096019",
    year = "2019"
}

@article{Mazumdar:2018dfl,
      author         = "Mazumdar, Anupam and White, Graham",
      title          = "{Review of cosmic phase transitions: their significance
                        and experimental signatures}",
      journal        = "Rept. Prog. Phys.",
      volume         = "82",
      year           = "2019",
      number         = "7",
      pages          = "076901",
      doi            = "10.1088/1361-6633/ab1f55",
      eprint         = "1811.01948",
      archivePrefix  = "arXiv",
      primaryClass   = "hep-ph",
      SLACcitation   = "%%CITATION = ARXIV:1811.01948;%%"
}

@article{Hindmarsh:2020hop,
    author = {Hindmarsh, Mark B. and L\"uben, Marvin and Lumma, Johannes and Pauly, Martin},
    title = "{Phase transitions in the early universe}",
    eprint = "2008.09136",
    archivePrefix = "arXiv",
    primaryClass = "astro-ph.CO",
    reportNumber = "MPP-2020-163, HIP-2020-27/TH",
    doi = "10.21468/SciPostPhysLectNotes.24",
    journal = "SciPost Phys. Lect. Notes",
    volume = "24",
    pages = "1",
    year = "2021"
}

@article{Caldwell:2022qsj,
    author = "Caldwell, Robert and others",
    title = "{Detection of early-universe gravitational-wave signatures and fundamental physics}",
    eprint = "2203.07972",
    archivePrefix = "arXiv",
    primaryClass = "gr-qc",
    doi = "10.1007/s10714-022-03027-x",
    journal = "Gen. Rel. Grav.",
    volume = "54",
    number = "12",
    pages = "156",
    year = "2022"
}

@article{Hawking:1982ga,
    author = "Hawking, S. W. and Moss, I. G. and Stewart, J. M.",
    title = "{Bubble Collisions in the Very Early Universe}",
    reportNumber = "Print-82-0180 (CAMBRIDGE)",
    doi = "10.1103/PhysRevD.26.2681",
    journal = "Phys. Rev. D",
    volume = "26",
    pages = "2681",
    year = "1982"
}

@article{Witten:1984rs,
    author = "Witten, Edward",
    title = "{Cosmic Separation of Phases}",
    reportNumber = "PRINT-84-0400 (IAS,PRINCETON)",
    doi = "10.1103/PhysRevD.30.272",
    journal = "Phys. Rev. D",
    volume = "30",
    pages = "272--285",
    year = "1984"
}

@article{Jinno:2016vai,
      author         = "Jinno, Ryusuke and Takimoto, Masahiro",
      title          = "{Gravitational waves from bubble collisions: An analytic
                        derivation}",
      journal        = "Phys. Rev.",
      volume         = "D95",
      year           = "2017",
      number         = "2",
      pages          = "024009",
      doi            = "10.1103/PhysRevD.95.024009",
      eprint         = "1605.01403",
      archivePrefix  = "arXiv",
      primaryClass   = "astro-ph.CO",
      reportNumber   = "KEK-TH-1900",
      SLACcitation   = "%%CITATION = ARXIV:1605.01403;%%"
}

@article{Hogan:1986dsh,
    author = "Hogan, C. J.",
    title = "{Gravitational radiation from cosmological phase transitions}",
    doi = "10.1093/mnras/218.4.629",
    journal = "Mon. Not. Roy. Astron. Soc.",
    volume = "218",
    number = "4",
    pages = "629--636",
    year = "1986"
}

@article{Hindmarsh:2016lnk,
      author         = "Hindmarsh, Mark",
      title          = "{Sound shell model for acoustic gravitational wave
                        production at a first-order phase transition in the early
                        Universe}",
      journal        = "Phys. Rev. Lett.",
      volume         = "120",
      year           = "2018",
      number         = "7",
      pages          = "071301",
      doi            = "10.1103/PhysRevLett.120.071301",
      eprint         = "1608.04735",
      archivePrefix  = "arXiv",
      primaryClass   = "astro-ph.CO",
      SLACcitation   = "%%CITATION = ARXIV:1608.04735;%%"
}

@article{Hindmarsh:2019phv,
      author         = "Hindmarsh, Mark and Hijazi, Mulham",
      title          = "{Gravitational waves from first order cosmological phase
                        transitions in the Sound Shell Model}",
      journal        = "JCAP",
      volume         = "1912",
      year           = "2019",
      pages          = "062",
      doi            = "10.1088/1475-7516/2019/12/062",
      eprint         = "1909.10040",
      archivePrefix  = "arXiv",
      primaryClass   = "astro-ph.CO",
      reportNumber   = "NORDITA-2019-083, HIP-2019-29/TH",
      SLACcitation   = "%%CITATION = ARXIV:1909.10040;%%"
}

@article{Cai:2023guc,
    author = "Cai, Rong-Gen and Wang, Shao-Jiang and Yuwen, Zi-Yan",
    title = "{Hydrodynamic sound shell model}",
    eprint = "2305.00074",
    archivePrefix = "arXiv",
    primaryClass = "gr-qc",
    doi = "10.1103/PhysRevD.108.L021502",
    journal = "Phys. Rev. D",
    volume = "108",
    number = "2",
    pages = "L021502",
    year = "2023"
}

@article{Kamionkowski:1993fg,
    author = "Kamionkowski, Marc and Kosowsky, Arthur and Turner, Michael S.",
    title = "{Gravitational radiation from first order phase transitions}",
    eprint = "astro-ph/9310044",
    archivePrefix = "arXiv",
    reportNumber = "IASSNS-HEP-93-44, FERMILAB-PUB-93-235-A",
    doi = "10.1103/PhysRevD.49.2837",
    journal = "Phys. Rev. D",
    volume = "49",
    pages = "2837--2851",
    year = "1994"
}

@article{Cai:2018teh,
      author         = "Cai, Rong-Gen and Wang, Shao-Jiang",
      title          = "{Energy budget of cosmological first-order phase
                        transition in FLRW background}",
      journal        = "Sci. China Phys. Mech. Astron.",
      volume         = "61",
      year           = "2018",
      pages          = "080411",
      doi            = "10.1007/s11433-018-9216-7",
      eprint         = "1803.03002",
      archivePrefix  = "arXiv",
      primaryClass   = "gr-qc",
      SLACcitation   = "%%CITATION = ARXIV:1803.03002;%%"
}

@article{Guo:2020grp,
    author = "Guo, Huai-Ke and Sinha, Kuver and Vagie, Daniel and White, Graham",
    title = "{Phase Transitions in an Expanding Universe: Stochastic Gravitational Waves in Standard and Non-Standard Histories}",
    eprint = "2007.08537",
    archivePrefix = "arXiv",
    primaryClass = "hep-ph",
    doi = "10.1088/1475-7516/2021/01/001",
    journal = "JCAP",
    volume = "01",
    pages = "001",
    year = "2021"
}

@article{Giombi:2025tkv,
    author = "Giombi, Lorenzo and Dahl, Jani and Hindmarsh, Mark",
    title = "{Acoustic gravitational waves beyond leading order in bubble over Hubble radius}",
    eprint = "2504.08037",
    archivePrefix = "arXiv",
    primaryClass = "gr-qc",
    reportNumber = "HIP-2025-13/TH",
    month = "4",
    year = "2025"
}

@article{Jinno:2023vnr,
    author = "Jinno, Ryusuke and Kume, Jun'ya and Yamada, Masaki",
    title = "{Super-slow phase transition catalyzed by BHs and the birth of baby BHs}",
    eprint = "2310.06901",
    archivePrefix = "arXiv",
    primaryClass = "hep-ph",
    reportNumber = "TU-1209, RESCEU-18/23",
    doi = "10.1016/j.physletb.2024.138465",
    journal = "Phys. Lett. B",
    volume = "849",
    pages = "138465",
    year = "2024"
}

@article{Yuwen:2024gcf,
    author = "Yuwen, Zi-Yan and Joana, Cristian and Wang, Shao-Jiang and Cai, Rong-Gen",
    title = "{Bubbles kick off primordial black holes to form more binaries}",
    eprint = "2406.05838",
    archivePrefix = "arXiv",
    primaryClass = "gr-qc",
    doi = "10.1103/PhysRevResearch.7.023180",
    journal = "Phys. Rev. Res.",
    volume = "7",
    number = "2",
    pages = "023180",
    year = "2025"
}

@article{Wang:2025hwc,
    author = "Wang, Haonan and Zhang, Ying-li and Suyama, Teruaki",
    title = "{Nearly Monochromatic Primordial Black Holes as total Dark Matter from Bubble Collapse}",
    eprint = "2510.19233",
    archivePrefix = "arXiv",
    primaryClass = "astro-ph.CO",
    month = "10",
    year = "2025"
}

@article{Devoto:2022qen,
    author = "Devoto, Federica and Devoto, Simone and Di Luzio, Luca and Ridolfi, Giovanni",
    title = "{False vacuum decay: an introductory review}",
    eprint = "2205.03140",
    archivePrefix = "arXiv",
    primaryClass = "hep-ph",
    doi = "10.1088/1361-6471/ac7f24",
    journal = "J. Phys. G",
    volume = "49",
    number = "10",
    pages = "103001",
    year = "2022"
}

@article{Moss:1984zf,
    author = "Moss, I. G.",
    title = "{BLACK HOLE BUBBLES}",
    reportNumber = "Print-85-0005 (NEWCASTLE)",
    doi = "10.1103/PhysRevD.32.1333",
    journal = "Phys. Rev. D",
    volume = "32",
    pages = "1333",
    year = "1985"
}

@article{Hiscock:1987hn,
    author = "Hiscock, W. A.",
    title = "{CAN BLACK HOLES NUCLEATE VACUUM PHASE TRANSITIONS?}",
    doi = "10.1103/PhysRevD.35.1161",
    journal = "Phys. Rev. D",
    volume = "35",
    pages = "1161--1170",
    year = "1987"
}

@article{Gregory:2013hja,
    author = "Gregory, Ruth and Moss, Ian G. and Withers, Benjamin",
    title = "{Black holes as bubble nucleation sites}",
    eprint = "1401.0017",
    archivePrefix = "arXiv",
    primaryClass = "hep-th",
    reportNumber = "DCPT-13-43",
    doi = "10.1007/JHEP03(2014)081",
    journal = "JHEP",
    volume = "03",
    pages = "081",
    year = "2014"
}

@article{Burda:2015isa,
    author = "Burda, Philipp and Gregory, Ruth and Moss, Ian",
    title = "{Gravity and the stability of the Higgs vacuum}",
    eprint = "1501.04937",
    archivePrefix = "arXiv",
    primaryClass = "hep-th",
    reportNumber = "DCPT-15-03",
    doi = "10.1103/PhysRevLett.115.071303",
    journal = "Phys. Rev. Lett.",
    volume = "115",
    pages = "071303",
    year = "2015"
}

@article{Burda:2015yfa,
    author = "Burda, Philipp and Gregory, Ruth and Moss, Ian",
    title = "{Vacuum metastability with black holes}",
    eprint = "1503.07331",
    archivePrefix = "arXiv",
    primaryClass = "hep-th",
    reportNumber = "DCPT-15-11",
    doi = "10.1007/JHEP08(2015)114",
    journal = "JHEP",
    volume = "08",
    pages = "114",
    year = "2015"
}

@article{Oshita:2019jan,
    author = "Oshita, Naritaka and Ueda, Kazushige and Yamaguchi, Masahide",
    title = "{Vacuum decays around spinning black holes}",
    eprint = "1909.01378",
    archivePrefix = "arXiv",
    primaryClass = "hep-th",
    doi = "10.1007/JHEP01(2020)015",
    journal = "JHEP",
    volume = "01",
    pages = "015",
    year = "2020",
    note = "[Erratum: JHEP 10, 122 (2020)]"
}

@article{Mukaida:2017bgd,
    author = "Mukaida, Kyohei and Yamada, Masaki",
    title = "{False Vacuum Decay Catalyzed by Black Holes}",
    eprint = "1706.04523",
    archivePrefix = "arXiv",
    primaryClass = "hep-th",
    reportNumber = "IPMU-17-0069",
    doi = "10.1103/PhysRevD.96.103514",
    journal = "Phys. Rev. D",
    volume = "96",
    number = "10",
    pages = "103514",
    year = "2017"
}

@article{Masoumi:2017trx,
    author = "Masoumi, Ali and Olum, Ken D. and Wachter, Jeremy M.",
    title = "{Approximating tunneling rates in multi-dimensional field spaces}",
    eprint = "1702.00356",
    archivePrefix = "arXiv",
    primaryClass = "gr-qc",
    doi = "10.1088/1475-7516/2017/10/022",
    journal = "JCAP",
    volume = "10",
    pages = "022",
    year = "2017",
    note = "[Erratum: JCAP 05, E01 (2023)]"
}

@article{Guada:2020xnz,
    author = "Guada, Victor and Nemev\v{s}ek, Miha and Pintar, Matev\v{z}",
    title = "{FindBounce: Package for multi-field bounce actions}",
    eprint = "2002.00881",
    archivePrefix = "arXiv",
    primaryClass = "hep-ph",
    doi = "10.1016/j.cpc.2020.107480",
    journal = "Comput. Phys. Commun.",
    volume = "256",
    pages = "107480",
    year = "2020"
}

@article{Wainwright:2011kj,
    author = "Wainwright, Carroll L.",
    title = "{CosmoTransitions: Computing Cosmological Phase Transition Temperatures and Bubble Profiles with Multiple Fields}",
    eprint = "1109.4189",
    archivePrefix = "arXiv",
    primaryClass = "hep-ph",
    doi = "10.1016/j.cpc.2012.04.004",
    journal = "Comput. Phys. Commun.",
    volume = "183",
    pages = "2006--2013",
    year = "2012"
}

@article{Shoji:2025nvj,
    author = "Shoji, Yutaro and Yamaguchi, Masahide",
    title = "{Symmetry of Bounce Solutions at Finite Temperature}",
    eprint = "2511.05950",
    archivePrefix = "arXiv",
    primaryClass = "hep-th",
    month = "11",
    year = "2025"
}

@article{Salvio:2016mvj,
    author = "Salvio, Alberto and Strumia, Alessandro and Tetradis, Nikolaos and Urbano, Alfredo",
    title = "{On gravitational and thermal corrections to vacuum decay}",
    eprint = "1608.02555",
    archivePrefix = "arXiv",
    primaryClass = "hep-ph",
    reportNumber = "CERN-TH-2016-180",
    doi = "10.1007/JHEP09(2016)054",
    journal = "JHEP",
    volume = "09",
    pages = "054",
    year = "2016"
}

@article{Vicentini:2022pra,
    author = "Vicentini, Silvia",
    title = "{New bounds on vacuum decay in de Sitter space}",
    eprint = "2205.11036",
    archivePrefix = "arXiv",
    primaryClass = "gr-qc",
    month = "5",
    year = "2022"
}

@article{Antoniadis:2024ent,
    author = "Antoniadis, Ignatios and Bielli, Daniele and Chatrabhuti, Auttakit and Isono, Hiroshi",
    title = "{Thin-wall vacuum decay in the presence of a compact dimension}",
    eprint = "2405.16920",
    archivePrefix = "arXiv",
    primaryClass = "hep-th",
    doi = "10.1007/JHEP09(2024)011",
    journal = "JHEP",
    volume = "09",
    pages = "011",
    year = "2024"
}

@article{Masoumi:2012yy,
    author = "Masoumi, Ali and Weinberg, Erick J.",
    title = "{Bounces with O(3) x O(2) symmetry}",
    eprint = "1207.3717",
    archivePrefix = "arXiv",
    primaryClass = "hep-th",
    doi = "10.1103/PhysRevD.86.104029",
    journal = "Phys. Rev. D",
    volume = "86",
    pages = "104029",
    year = "2012"
}

@article{Lee:2021nwg,
    author = "Lee, Bum-Hoon and Lee, Wonwoo and Yeom, Dong-han and Yin, Lu",
    title = "{Gravitational waves from the vacuum decay with LISA *}",
    eprint = "2106.07430",
    archivePrefix = "arXiv",
    primaryClass = "gr-qc",
    doi = "10.1088/1674-1137/ac5d2a",
    journal = "Chin. Phys. C",
    volume = "46",
    number = "7",
    pages = "075101",
    year = "2022"
}

@article{Coleman:1980aw,
    author = "Coleman, Sidney R. and De Luccia, Frank",
    title = "{Gravitational Effects on and of Vacuum Decay}",
    reportNumber = "SLAC-PUB-2463",
    doi = "10.1103/PhysRevD.21.3305",
    journal = "Phys. Rev. D",
    volume = "21",
    pages = "3305",
    year = "1980"
}

@article{Hawking:1981fz,
    author = "Hawking, S. W. and Moss, I. G.",
    title = "{Supercooled Phase Transitions in the Very Early Universe}",
    reportNumber = "Print-82-0181 (CAMBRIDGE)",
    doi = "10.1016/0370-2693(82)90946-7",
    journal = "Phys. Lett. B",
    volume = "110",
    pages = "35--38",
    year = "1982"
}

@article{Shibata:1999zs,
    author = "Shibata, Masaru and Sasaki, Misao",
    title = "{Black hole formation in the Friedmann universe: Formulation and computation in numerical relativity}",
    eprint = "gr-qc/9905064",
    archivePrefix = "arXiv",
    reportNumber = "OU-TAP-93",
    doi = "10.1103/PhysRevD.60.084002",
    journal = "Phys. Rev. D",
    volume = "60",
    pages = "084002",
    year = "1999"
}

@article{Joana:2025gqf,
    author = "Joana, Cristian and Yuwen, Zi-Yan",
    title = "{Primordial Black Holes from Primordial Voids}",
    eprint = "2510.11611",
    archivePrefix = "arXiv",
    primaryClass = "astro-ph.CO",
    month = "10",
    year = "2025"
}

@article{Bardeen:1985tr,
    author = "Bardeen, James M. and Bond, J. R. and Kaiser, Nick and Szalay, A. S.",
    title = "{The Statistics of Peaks of Gaussian Random Fields}",
    reportNumber = "FERMILAB-PUB-85-148-A, NSF-ITP-85-93",
    doi = "10.1086/164143",
    journal = "Astrophys. J.",
    volume = "304",
    pages = "15--61",
    year = "1986"
}

@article{Yoo:2018kvb,
    author = "Yoo, Chul-Moon and Harada, Tomohiro and Garriga, Jaume and Kohri, Kazunori",
    title = "{Primordial black hole abundance from random Gaussian curvature perturbations and a local density threshold}",
    eprint = "1805.03946",
    archivePrefix = "arXiv",
    primaryClass = "astro-ph.CO",
    reportNumber = "RUP-18-15, KEK-Cosmo-225, KEK-TH-2052",
    doi = "10.1093/ptep/pty120",
    journal = "PTEP",
    volume = "2018",
    number = "12",
    pages = "123E01",
    year = "2018",
    note = "[Erratum: PTEP 2024, 049202 (2024)]"
}

@article{Yoo:2020dkz,
    author = "Yoo, Chul-Moon and Harada, Tomohiro and Hirano, Shin'ichi and Kohri, Kazunori",
    title = "{Abundance of Primordial Black Holes in Peak Theory for an Arbitrary Power Spectrum}",
    eprint = "2008.02425",
    archivePrefix = "arXiv",
    primaryClass = "astro-ph.CO",
    reportNumber = "RUP-20-25, KEK-Cosmo-261, KEK-TH-2245",
    doi = "10.1093/ptep/ptaa155",
    journal = "PTEP",
    volume = "2021",
    number = "1",
    pages = "013E02",
    year = "2021",
    note = "[Erratum: PTEP 2024, 049203 (2024)]"
}

@article{Musco:2018rwt,
    author = "Musco, Ilia",
    title = "{Threshold for primordial black holes: Dependence on the shape of the cosmological perturbations}",
    eprint = "1809.02127",
    archivePrefix = "arXiv",
    primaryClass = "gr-qc",
    doi = "10.1103/PhysRevD.100.123524",
    journal = "Phys. Rev. D",
    volume = "100",
    number = "12",
    pages = "123524",
    year = "2019"
}

@article{Lewicki:2019gmv,
    author = "Lewicki, Marek and Vaskonen, Ville",
    title = "{On bubble collisions in strongly supercooled phase transitions}",
    eprint = "1912.00997",
    archivePrefix = "arXiv",
    primaryClass = "astro-ph.CO",
    reportNumber = "KCL-PH-TH/2019-88",
    doi = "10.1016/j.dark.2020.100672",
    journal = "Phys. Dark Univ.",
    volume = "30",
    pages = "100672",
    year = "2020"
}

@article{Coleman:1977py,
    author = "Coleman, Sidney R.",
    title = "{The Fate of the False Vacuum. 1. Semiclassical Theory}",
    reportNumber = "HUTP-77-A004",
    doi = "10.1103/PhysRevD.16.1248",
    journal = "Phys. Rev. D",
    volume = "15",
    pages = "2929--2936",
    year = "1977",
    note = "[Erratum: Phys.Rev.D 16, 1248 (1977)]"
}

@article{Callan:1977pt,
    author = "Callan, Jr., Curtis G. and Coleman, Sidney R.",
    title = "{The Fate of the False Vacuum. 2. First Quantum Corrections}",
    reportNumber = "HUTP-77-A032",
    doi = "10.1103/PhysRevD.16.1762",
    journal = "Phys. Rev. D",
    volume = "16",
    pages = "1762--1768",
    year = "1977"
}

@article{Coleman:1977th,
    author = "Coleman, Sidney R. and Glaser, V. and Martin, Andre",
    title = "{Action Minima Among Solutions to a Class of Euclidean Scalar Field Equations}",
    reportNumber = "CERN-TH-2364",
    doi = "10.1007/BF01609421",
    journal = "Commun. Math. Phys.",
    volume = "58",
    pages = "211--221",
    year = "1978"
}

@article{Linde:1980tt,
    author = "Linde, Andrei D.",
    title = "{Fate of the False Vacuum at Finite Temperature: Theory and Applications}",
    reportNumber = "LEBEDEV-80-92",
    doi = "10.1016/0370-2693(81)90281-1",
    journal = "Phys. Lett. B",
    volume = "100",
    pages = "37--40",
    year = "1981"
}

@article{Linde:1981zj,
    author = "Linde, Andrei D.",
    title = "{Decay of the False Vacuum at Finite Temperature}",
    reportNumber = "LEBEDEV-81-265",
    doi = "10.1016/0550-3213(83)90293-6",
    journal = "Nucl. Phys. B",
    volume = "216",
    pages = "421",
    year = "1983",
    note = "[Erratum: Nucl.Phys.B 223, 544 (1983)]"
}

@article{Guth:1982pn,
    author = "Guth, Alan H. and Weinberg, Erick J.",
    title = "{Could the Universe Have Recovered from a Slow First Order Phase Transition?}",
    reportNumber = "MIT-CTP-950",
    doi = "10.1016/0550-3213(83)90307-3",
    journal = "Nucl. Phys. B",
    volume = "212",
    pages = "321--364",
    year = "1983"
}

@article{Turner:1992tz,
    author = "Turner, Michael S. and Weinberg, Erick J. and Widrow, Lawrence M.",
    title = "{Bubble nucleation in first order inflation and other cosmological phase transitions}",
    reportNumber = "FERMILAB-PUB-91-334-A, CU-TP-558, IASSNS-HEP-92-21",
    doi = "10.1103/PhysRevD.46.2384",
    journal = "Phys. Rev. D",
    volume = "46",
    pages = "2384--2403",
    year = "1992"
}

@book{Calzetta:2008iqa,
    author = "Calzetta, Esteban A. and Hu, Bei-Lok B.",
    title = "{Nonequilibrium Quantum Field Theory}",
    doi = "10.1017/9781009290036",
    isbn = "978-1-009-29003-6, 978-1-009-28998-6, 978-1-009-29002-9, 978-0-511-42147-1, 978-0-521-64168-5",
    publisher = "Oxford University Press",
    year = "2009"
}

@article{Jinno:2024nwb,
    author = "Jinno, Ryusuke and Kume, Jun'ya",
    title = "{Gravitational effects on fluid dynamics in cosmological first-order phase transitions}",
    eprint = "2408.10770",
    archivePrefix = "arXiv",
    primaryClass = "gr-qc",
    reportNumber = "RESCEU-12/24, KOBE-COSMO-24-02",
    doi = "10.1088/1475-7516/2025/02/057",
    journal = "JCAP",
    volume = "02",
    pages = "057",
    year = "2025"
}

\end{document}